\documentclass[final,authoryear,3p,11pt]{elsarticle}

\usepackage{natbib} 
\usepackage{booktabs}
\usepackage{color}
\usepackage{soul}
\usepackage{lscape}
\sethlcolor{red}
\usepackage{hyperref}
\usepackage{graphicx}
\usepackage{amssymb}
\usepackage{amsfonts}
\usepackage{amsmath}
\usepackage{epstopdf}
\usepackage[font={footnotesize}]{caption}
\usepackage{acronym}
  \acrodef{HAM}{Heterogeneous Agent Model}
  \acrodef{CT}{Catastrophe Theory}
  \acrodef{RV}{Realized Volatility}

\begin{document}

\title{Realizing stock market crashes: stochastic cusp catastrophe model of returns under the time-varying volatility\tnoteref{label1}}
\author[ies,utia]{Jozef Barunik\corref{cor2}} \ead{barunik@utia.cas.cz}
\author[ies,utia]{Jiri Kukacka} \ead{jiri.kukacka@gmail.com}
\cortext[cor2]{Corresponding author}
\address[ies]{Institute of Economic Studies, Faculty of Social Sciences, Charles University, Opletalova 21, 110 00, Prague, Czech Republic}
\address[utia]{Institute of Information Theory and Automation, Academy of Sciences of the Czech Republic, Pod Vodarenskou vezi 4, 182 00, Prague, Czech Republic}
\tnotetext[label1]{The support from the Czech Science Foundation under the 402/09/0965, 13-32263S and 13-24313S projects and Grant Agency of the Charles University under the 588912 project is gratefully acknowledged.}

\begin{abstract}
This paper develops a two-step estimation methodology, which allows us to apply catastrophe theory to stock market returns with time-varying volatility and model stock market crashes. Utilizing high frequency data, we estimate the daily realized volatility from the returns in the first step and use stochastic cusp catastrophe on data normalized by the estimated volatility in the second step to study possible discontinuities in markets. We support our methodology by simulations where we also discuss the importance of stochastic noise and volatility in deterministic cusp catastrophe model. The methodology is empirically tested on almost 27 years of U.S. stock market evolution covering several important recessions and crisis periods. Due to the very long sample period we also develop a rolling estimation approach and we find that while in the first half of the period stock markets showed marks of bifurcations, in the second half catastrophe theory was not able to confirm this behavior. Results suggest that the proposed methodology provides an important shift in application of catastrophe theory to stock markets.
\end{abstract}
\maketitle


\section{Introduction}

Financial inefficiencies such as under- or over-reactions to information as causes of extreme events in the stock markets are attracting researchers across all fields of economics. 
In one of the recent contributions, \citet{Levy2008} highlights the endogeneity of large market crashes as a result of a natural conformity of investors with their peers and the heterogeneity of investors' population. The stronger the conformity and homogeneity across the market, the more likely is the existence of multiple equilibria in the market, which is a prerequisite for a market crash to happen. \citet{GeLe1990} present a model sharing the same notions as \citet{Levy2008} in terms of effect of small changes when market is close the crash point as well as the volatility amplification signaling. In another work, \citet{LeLeSo1994} take into consideration the signals produced by dividend yields and assess the effect of computer trading being blamed for making the market more homogeneous and thus contributive to uprise of a crash. \citet{Kleidon1995} summarizes and compares several older models from the turn of 1980s and 1990s, and \citet{BaVe2003} propose a model based on rational but uninformed traders, who can unreasonably panic. In this approach, again, abrupt declines in stock prices can happen even without any real change of underlying fundamentals. \citet{Lux1995} links the phenomena of market crashes to the process of phase transition from thermodynamics and models the emergence of bubbles and crashes as a result of herd behavior among heterogeneous traders in speculative markets. Finally, a strand of literature documenting precursory patterns and log-periodic signatures even years before largest crashes in the modern history advocates that crash has an endogenous origin in the `crowd' behavior and interactions of many agents \citep{SoJo1998b,JoLeSo2000,Sornette2002,Sornette2004}. Compared to commonly shared beliefs, Didier Sornette and his colleagues argue that exogenous shocks can only serve as triggers, not as the direct causes of crashes and that large crashes are `outliers'.

A very different approach attempting to unfold the information we might need to understand the crash phenomena by describing how even small shifts in the speculative part of the market can have sudden, discontinuous effect on prices is catastrophe theory. Catastrophe theory has been proposed by French mathematician \citet{Thom1975} with the primal aim to shed some light on the `mystery' of biological morphogenesis. Despite its mathematical virtue, the theory has been promptly heavily criticized by \citet{ZaSu1977} and \citet{SuZa1978a,SuZa1978b} for its excessive utilization of qualitative approaches together with improper usage of statistical methods as well as for violation of the necessary mathematical assumptions in many applications. Due to this criticism, the intellectual bubble and the heyday of cusp catastrophe approach from 1970s downgraded rapidly, even if defended e.g. by \citet{Boutot1993} or the extensive, gradually updated work of \citet{Arnold2004}. Nonetheless, until quite recently the `fatal' criticism has been ridiculed by \citet[p.~3275 \& 3257]{Rosser2007}, who states that \textit{``the baby of catastrophe theory was largely thrown out with the bathwater of its inappropriate applications''} and suggests that \textit{``economists should reevaluate the former fad and move it to a more proper valuation''}.

The application of catastrophe theory in social sciences is not as wide as in natural sciences, however, it has been proposed from its early birth. \cite{zeeman1974}'s cooperation with \citeauthor{Thom1975} and also popularization of the theory by his own nontechnical examples \citep{Zeeman1975,Zeeman1976} led to the development of many applications in the fields of economics, psychology, sociology, political studies, and others. \citet{zeeman1974} also proposed the application of cusp catastrophe model to stock markets. Translating seven qualitative hypotheses of stock market exchanges to mathematical terminology of catastrophe theory, one of the first heterogeneous agent models of two main types of investors: fundamentalists and chartists, came into the world.
Heterogeneity and interaction between the two distinct types of agents have attracted wider behavioral finance literature. Fundamentalists base their expectations about future asset prices in their beliefs about fundamental and economic factors such as dividends, earnings, macroeconomic environment. In contrast, chartists do not consider fundamentals in their trading strategies at all. Their expectations about future asset prices are based on finding historical patterns in prices. 
While Zeeman's work was only qualitative description of the observed facts as bull and bear markets, it contains a number of important behavioral elements that have been used later in large amount of literature focusing on heterogeneous agent modeling.\footnote{For a recent survey of heterogeneous agent models, see \cite{Hommes2006}. Special issue on heterogeneous interacting agents in financial markets edited by \cite{lux2002} also provides an interesting contributions.}

Nowadays the statistical theory is well developed and parametrized cusp catastrophe models can be evaluated quantitatively on the data. 
The biggest difficulty in application of catastrophe theory arises from the fact that it stems from deterministic systems. Thus, it is hard to apply it directly to systems that are subject to random influences which are common in behavioral sciences. \citet{cobb1980,Cobb1981} and \citet{cobb1985} provided the bridge and took catastrophe theory from determinism to stochastic systems. While this was an important shift, there are further complications in application to stock markets data. Main restriction of Cobb's methodology of estimation is the requirement for constant variance which directly forces us to assume that volatility of the stock markets (as standard deviation of the returns) is constant. Quantitative verification of \cite{zeeman1974}'s hypotheses about application of the theory to stock market crashes has been pioneered by \cite{barunik2009}, where we fit the cusp model to the two separate events of large known stock market crashes. Although a successful application, \cite{barunik2009} bring only preliminary results in the restricted environment. Application of cusp catastrophe theory on stock market data deserves lot more attention. In the current paper, we propose an improved way of application and we believe to bring closer answer to the question whether cusp catastrophe theory is capable of explaining stock market crashes. 

Time-varying volatility has become an important stylized fact about stock market data and researchers quickly recognized that it is an important feature of any modeling strategy used. One of the most successful early works of \cite{engle,bollerslev} proposed to include the volatility as a time-varying process in the (generalized) autoregressive conditional heteroskedasticity framework. From these times, many models has been developed in the literature which improved the original frameworks. As early as in `late nineties', high frequency data became available to researchers and this has led to another important shift in volatility modeling --- realized volatility. A theory for very simple and intuitive approach of computing the daily volatility using sum of squared high-frequency returns has been formalized by \cite{abdl2003,barndorff2004b}. While the volatility literature is immense,\footnote{\cite{andersen2004parametric} provides a very useful and complete review of the methodologies.} several researchers have also studied the volatility and stock market crashes. For example, \cite{shaffer} argues that volatility might be the cause of a stock market crash. In contrast, \cite{Levy2008} explains why volatility increases before the crash even if no dramatic information has been revealed.

In our work, we utilize the availability of the high frequency data and popular realized volatility approach to propose a two-step method of estimation, which will overcome the difficulties in application of cusp catastrophe theory on stock market data. Using realized volatility, we estimate the stock market returns volatility and then apply stochastic cusp catastrophe model on volatility adjusted returns with constant variance. This is motivated by confirmed bimodal distributions of such standardized data at some periods and it allows us to study if stock markets are driven into catastrophe endogenously or whether it is just an effect of volatility. To motivate our approach, we also run a simulation study which gives us strong support for the methodology. Simulations also illustrate the importance of stochastic noise and volatility in the deterministic cusp model.

Using a unique data covering almost 27 years of the U.S. stock markets evolution, we empirically test stochastic cusp catastrophe model in the time-varying volatility environment. Moreover, we take advantage of the data and develop a rolling regression approach to study the dynamics of the model parameters in such a long time covering several important recessions and crisis. This allows us to localize the bifurcation periods. 

We need to mention several important works which bring similar results to ours. \cite{creedy1993multiple,creedy1996non} develop a framework for estimation of non-linear exchange rate models and they show that swings in the exchange rates can be attributed to the bimodality even without explicit use of catastrophe theory. More recently, \cite{koh2007cardan} propose to use the Cardan's dicriminant to detect the bimodality and confirm the predictive ability of currency pairs for emerging countries. In our work, we bring new insight to the non-linear phenomena allowing for time-varying volatility into the modelling strategy.  

The paper is organized as follows. The second and the third Sections bring theoretical framework of stochastic catastrophe theory under the time-varying volatility and its estimation. The fourth Section presents the simulations supporting our two-step method of estimation and is followed by the empirical application of the theory on the stock market crashes modeling. Finally, last Section concludes. 


\section{Theoretical framework}

Catastrophe theory has been developed as a deterministic theory for systems that may respond to continuous changes in control variables by a discontinuous change from one equilibrium state to another. A key idea is that system under study is driven towards an equilibrium state. The behavior of the dynamical systems under study is completely determined by a so-called potential function, which depends on behavioral and control variables. The behavioral, or state variable describes the state of the system, while control variables determine the behavior of the system. The dynamics under catastrophe models can become extremely complex, and according to the classification theory of \cite{Thom1975}, there are seven different families based on the number of control and dependent variables. We focus on the application of catastrophe theory to model sudden stock market crashes as qualitatively proposed by \cite{zeeman1974}. In his work, Zeeman used the so-called cusp catastrophe model, the simplest specification that gives rise to sudden discontinuities. 

\subsection{Deterministic dynamics}

Let us suppose that the process $y_t$ evolves over $t=1,\ldots,T$ as
\begin{equation}
\label{eq:y}
d y_t=-\frac{dV(y_t;\alpha,\beta)}{dy_t}dt,
\end{equation}
where $V(y_t;\alpha,\beta)$ is the potential function describing the dynamics of the state variable $y_t$ controlled by parameters $\alpha$ and $\beta$ determining the system. When the right-hand side of Eq. (\ref{eq:y}) equals zero, $-dV(y_t;\alpha,\beta)/dy_t=0$, the system is in equilibrium. If the system is at a non-equilibrium point, it will move back to its equilibrium where the potential function takes the minimum values with respect to $y_t$. While the concept of potential function is very general, i.e. it can be quadratic yielding equilibrium of a simple flat response surface, one of the most applied potential functions in behavioral sciences, a cusp potential function is defined as 
\begin{equation}
\label{eq:cuspdet}
-V(y_t;\alpha,\beta)=-1/4 y_t^4+1/2\beta y_t^2+\alpha y_t,
\end{equation}
with equilibria at 
\begin{equation}
\label{eq:cuspeq}
-\frac{dV(y_t;\alpha,\beta)}{dy_t}=- y_t^3+\beta y_t+\alpha
\end{equation}
being equal to zero. The two dimensions of the control space, $\alpha$ and $\beta$, further depend on realizations from $i=1\ldots,n$ independent variables $x_{i,t}$. Thus it is convenient to think about them as functions
\begin{eqnarray}
\alpha_x&=&\alpha_0+\alpha_1 x_{1,t}+\ldots+\alpha_n x_{n,t} \\
\beta_x&=&\beta_0+\beta_1 x_{1,t}+\ldots+\beta_n x_{n,t}.
\end{eqnarray}
The control functions $\alpha_x$ and $\beta_x$ are called normal and splitting factors, or asymmetry and bifurcation factors, respectively \citep{stewart} and they determine the predicted values of $y_t$ given $x_{i,t}$. This means that for each combination of values of independent variables there might be up to three predicted values of the state variable given by roots of
\begin{equation}
\label{eq:cusroots}
-\frac{dV(y_t;\alpha_x,\beta_x)}{dy_t}=- y_t^3+\beta_x y_t+\alpha_x=0 .
\end{equation}
This equation has one solution if 
\begin{equation}
\label{eq:cardan}
\delta_x=1/4\alpha_x^2-1/27\beta_x^3
\end{equation}
is greater than zero, $\delta_x>0$ and three solutions if $\delta_x<0$. This construction was first described by the $16^{th}$ century mathematician Geronimo Cardan and can serve as a statistic for bimodality, one of the catastrophe flags. The set of values for which Cardan's discriminant is equal to zero, $\delta_x=0$ is the bifurcation set which determines the set of singularity points in the system. In the case of three roots, the central root is called an ``anti-prediction" and is least probable state of the system. Inside the bifurcation, when $\delta_x<0$, the surface predicts two possible values of the state variable which means that the state variable is bimodal in this case.  For the illustration of the deterministic response surface of cusp catastrophe, we borrow the Figure \ref{fig:simulatedcusp} from simulations Section, where the deterministic response surface is a smooth pleat.

\subsection{Stochastic dynamics}

Most of the systems in behavioral sciences are subject to noise stemming from measurement errors or inherent stochastic nature of the system under study. Thus for a real-world applications, it is necessary to add non-deterministic behavior into the system. As catastrophe theory has primarily been developed to describe deterministic systems, it may not be obvious how to extend the theory to stochastic systems. An important bridge has been provided by \cite{cobb1980,Cobb1981} and \citet{cobb1985} who used the It\^{o} stochastic differential equations  to establish a link between the potential function of a deterministic catastrophe system and the stationary probability density function of the corresponding stochastic process. This in turn led to definition of a stochastic equilibrium state and bifurcation compatible with the deterministic counterpart. Cobb and his colleagues simply added a stochastic Gaussian white noise term to the system
\begin{equation}
\label{eq:ystoch}
d y_t=-\frac{dV(y_t;\alpha_x,\beta_x)}{dy_t}dt+\sigma_{y_t}dW_t,
\end{equation}
where $-dV(y_t;\alpha_x,\beta_x)/dy_t$ is the deterministic term, or drift function representing the equilibrium state of the cusp catastrophe, $\sigma_{y_t}$ is the diffusion function and $W_t$ is a Wiener process. When the diffusion function is constant, $\sigma_{y_t}=\sigma$, and the current measurement scale is not to be nonlinearly transformed, the stochastic potential function is proportional to deterministic potential function and probability distribution function corresponding to the solution from Eq. (\ref{eq:ystoch}) converges to a probability distribution function of a limiting stationary stochastic process as dynamics of $y_t$ are assumed to be much faster than changes in $x_{i,t}$ \citep{Cobb1981,cobb1985,wagenmakers2005}. The probability density that describes the distribution of the system's states at any $t$ is then
\begin{equation}
\label{eq:pdf}
f_s(y|x)=\psi \exp\left(\frac{(-1/4) y^4+(\beta_x/2) y^2+\alpha_x y}{\sigma} \right).
\end{equation}
The constant $\psi$ normalizes the probability distribution function so its integral over the entire range equals to one. As bifurcation factor $\beta_x$ changes from negative to positive, the $f_s(y|x)$ changes its shape from unimodal to bimodal. On the other hand, $\alpha_x$  causes asymmetry in $f_s(y|x)$.

\subsection{Cusp catastrophe under the time-varying volatility}

Stochastic catastrophe theory works only under the assumption that the diffusion function is constant, $\sigma_{y_t}=\sigma$, and the current measurement scale is not to be nonlinearly transformed. While this may be reliable in some applications in behavioral sciences, it may cause crucial difficulties in others. One of the problematic application is in modeling stock market crashes as the diffusion function $\sigma$, called volatility of stock market returns, has strong time-varying dynamics and it clusters over time, which is documented by strong dependence in the squared returns. To illustrate the volatility dynamics, let us borrow the dataset used further in this study. Figure \ref{fig:returns} shows the evolution of the S\&P 500 stock index returns over almost 27 years and documents how volatility strongly varies over time. One of the possible and very simple solutions in applying cusp catastrophe theory to the stock makrets is to consider only a short time window and fit the catastrophe model to the data where volatility can be assumed as constant \citep{barunik2009}. While in \cite{barunik2009} we were the first who quantitatively applied stochastic catastrophes to explain the stock market crashes on localized periods of crashes, this is very restrictive assumption to make generally.

Here, we propose more rigorous solution to the problem by utilizing recently developed concept of realized volatility. It will allow us to use the previously introduced concepts as after estimating volatility from the returns process consistently, we will be able to estimate the catastrophe model on the process which fulfils assumptions of stochastic catastrophe theory. Thus we assume that stock markets can be described by cusp catastrophe process subject to time-varying volatility. While this is a great advantage, which will allow us to apply cusp catastrophe theory on the different time periods conveniently, the disadvantage of our approach is that it can not be generalized to other branches of behavioral sciences where the high frequency data are not available and thus realized volatility can not be computed. Thus our generalization is restricted to application on financial data, mainly. Still, our main aim is to study the stock market crashes, thus the advocated approach will be very useful in the field of behavioral finance. Let us describe the theoretical concept here and in the next Sections, we will describe the full model and two-step estimation procedure.

Suppose that the sample path of the corresponding (latent) logarithmic price process $p_t$ is continuous over $t=1,\ldots,T$ and determined by the stochastic differential equation
\begin{equation} 
\label{eq:smme}
d p_t=\mu_tdt+\sigma_tdW_t,
\end{equation}
where $\mu_t$ is a drift term, $\sigma_t$ is the predictable diffusion function, or instantaneous volatility, and $W_t$ is a standard Brownian motion. A natural measure of the ex-post return variability over the $[t-h,t]$ time interval, $0 \le h \le t \le T$ is the integrated variance
\begin{equation}
IV_{t,h}=\int_{t-h}^{t}\sigma^2_{\tau}d\tau,
\end{equation}
which is not directly observed, but as shown in \cite{abdl2003} and \cite{barndorff2004b}, corresponding realized volatilities provide its consistent estimates. While it is convenient to work in the continuous time environment, empirical investigations are based on discretely sampled prices and we are interested in studying $h$-period continuously compounded discrete-time returns $r_{t,h}=p_t-p_{t-h}$. \cite{abdl2003} and \cite{barndorff2004b} showed that daily returns are Gaussian conditionally on information set generated by the sample paths of $\mu_t$ and $\sigma_t$ and integrated volatility normalizes the returns as 
\begin{equation}
r_{t,h} \left( \int_{t-h}^t\sigma^2_{\tau} d\tau \right)^{-1/2} \sim N\left(\int_{t-h}^t\mu_{\tau} d\tau,1\right).
\end{equation}
This result of quadratic variation theory is important to us as we use it to study stochastic cusp catastrophe in the environment where volatility is time-varying. In the modern stochastic volatility literature it is common to assume that stock market returns follow the very general semi-martingale process (as in Eq. \ref{eq:smme}), where the drift and volatility functions are predictable and the rest is unpredictable. Back in the origins of this literature, one of the very first contributions published in the stochastic volatility literature by \cite{taylor} assumed that daily returns will be the product of volatility and autoregression process. In our application, this will be also the case as we will assume that daily stock market returns are described by the process which is the product of volatility and cusp catastrophe model.

To formulate the approach, we assume that stock returns normalized by its volatility 
\begin{equation}
y_t^*=r_t \left( \int_{t-h}^t\sigma^2_{\tau} d\tau \right)^{-1/2} 
\end{equation}
follow a stochastic cusp catastrophe process
\begin{equation}
\label{eq:ystochnorm}
d y_t^*=-\frac{dV(y_t^*;\alpha_x,\beta_x)}{dy_t^*}dt+ dW_t.
\end{equation}
It is important to note the difference between Equation (\ref{eq:ystochnorm}) and Equation (\ref{eq:ystoch}) as there is no diffusion term in the process anymore. As the diffusion term of $y_t^*$ is constant and equal to one now, results of Cobb can be used conveniently, and we can use stationary probability distribution function of $y_t^*$ for the parameter estimation using maximum likelihood method.

As noted earlier, the integrated volatility is not directly observable. However, the now so popular concept of realized volatility and availability of high-frequency data brought a simple ways how to measure it accurately. This will help us in proposing a simple and intuitive way of estimating cusp catastrophe model on the stock market returns under the highly dynamic volatility.

\section{Estimation}

A simple consistent estimator of the integrated variance under the assumption of no microstructure noise contamination in the price process is provided by the well-known realized variance \citep{abdl2003,barndorff2004b} . The realized variance over $[t-h,t]$, for $0\le h \le t\le T$, is defined by
\begin{equation}
\label{rv}
\widehat{RV}_{t,h}=\sum_{i=1}^N r_{t-h+\left(\frac{i}{N}\right)h}^2,
\end{equation}
where $N$ is the number of observations in $[t-h,t]$ and $r_{t-h+\left(\frac{i}{N}\right)h}$ is $i-$th intraday return in the $[t-h,t]$ interval. $\widehat{RV}_{t}$ converges in probability to true integrated variance of the process as $N\rightarrow\infty$. 
\begin{equation}
\label{rv}
\widehat{RV}_{t,h} \overset{p}{\rightarrow} \int_{t-h}^t\sigma^2_{\tau} d\tau,
\end{equation}

As observed log-prices are contamined with microstructure noise in the real world, subsequent literature has developed several estimators. While it is important to consider both jumps and microstructure noise in the data, our main interest is in estimating catastrophe theory and addressing the question if it can be used to explain the deterministic part of stock market returns. Thus, we will restrict ourselves to the simplest estimator which uses sparse sampling to deal with the microstructure noise. Extant literature shows the support of this simple estimator, most recently, \cite{patton} run a horse race of most popular estimators and conclude that when the simple realized volatility is computed using 5-minute sampling, it is very hard to be outperformed.

In the first step, we thus estimate the realized volatility from the stock market returns using 5-min data as proposed by the theory and normalize the daily returns to obtain returns with constant volatility. While using the daily returns, $h=1$ and we drop $h$ for the ease of notation from now on.
\begin{equation}
\label{eq:rv}
\tilde{r}_t=r_t \widehat{RV}_{t}^{-1/2}
\end{equation}
In the second step, we apply stochastic cusp catastrophe to model the normalized stock market returns. While state variable of the cusp is a canonical variable, it is an unknown smooth transformation of the actual state variable of the system. As proposed by \cite{grasman}, we allow for the first order approximation to the true smooth transition allowing the measured $\tilde{r}$ to be a
\begin{equation}
\label{eq:esttransform}
y_t=\omega_0+\omega_1 \tilde{r}_t,
\end{equation}
with $\omega_1$ the first order coefficient of a polynomial approximation.
Independent variables will be
\begin{eqnarray}
\alpha_x&=&\alpha_0+\alpha_1 x_{1,t}+\ldots+\alpha_n x_{n,t} \\
\beta_x&=&\beta_0+\beta_1 x_{1,t}+\ldots+\beta_n x_{n,t},
\end{eqnarray}
hence the statistical estimation problem is to estimate $2n+2$ parameters $\{\omega_0,\omega_1,\alpha_0,\ldots,\alpha_n,\beta_0,\ldots,\ldots,\beta_n\}$. We estimate the parameters using the maximum likelihood approach  of \cite{cobb1980} augmented by \cite{grasman}. The negative log-likelihood for a sample of observed values $(x_{1,t},\ldots,x_{n,t},y_{t})$ for $t=1,\ldots,T$ is simply logarithm of the probability distribution function in Eq. (\ref{eq:pdf}). 

\subsection{Statistical evaluation of the fit}

To assess the fit of cusp catastrophe model to the data, a number of diagnostic tools have been suggested. \cite{stewart} suggested a pseudo-$R^2$ as a measure of explained variance. Difficulty arises here as for a given set of values of the independent variables the model may predict multiple values for the dependent variable. As a result of bimodal density, expected value is unlikely to be observed as it is an unstable solution in the equilibrium. For this reason, two alternatives for the expected value as the predictive value can be used. The first one chooses the mode of the density closest to the state values and is known as the delay convention, the second one uses the mode at which the density is highest and is known as the Maxwell convention. \cite{cobb1980} and \cite{stewart} suggest to use delay convention where the variance of error is defined as the variance of the difference between the estimated states and the mode of the distribution that is closest to this value. The pseudo-$R^2$ is then defined as $1-Var(\epsilon)/Var(y)$, where $\epsilon$ is error. 

While pseudo-$R^2$ is problematic due to the nature of cusp catastrophe model, it should be used complementary to other alternatives. To test the statistical fit of cusp catastrophe model rigorously, we use following steps. First, the cusp fit should be substantially better than multiple linear regression. This could be tested by means of the likelihood ratio test, which is asymptotically chi-squared distributed with degrees of freedom being equal to the difference in degrees of freedom for two compared models. Second, $\omega_1$ coefficient  should deviate significantly from zero. Otherwise, the $y_t$ in Eq. (\ref{eq:esttransform}) would be constant and cusp model would not describe the data. Third, the cusp model should indicate a better fit than following logistic curve
\begin{equation}
y_t=\frac{1}{1+e^{-\alpha_t/\beta_t^2}}+\epsilon_t,
\end{equation} 
for $t=1,\ldots,T$, where $\epsilon_t$ are zero mean random disturbances. The rationale for choosing to compare the cusp model to this logistic curve is that this function does not posses degenerate points, while it possibly models steep changes in the state variable as a function of changes in independent variables mimicking the sudden transitions of the cusp. Thus comparison of the cusp catastrophe model to the logistic function serves us as a good indicator of the presence of bifurcations in the data. While these two models are not nested, \cite{wagenmakers2005} suggests to compare them via information criteria, when stronger Bayesian Information Criterion (BIC) should be required for the decision. 


\section{Monte Carlo study}

In order to validate our assumptions about process generating stock market returns and our two-step estimation procedure, we conduct a Monte Carlo study where we simulate the data from stochastic cusp catastrophe model, allow for time-varying volatility in the process and estimate the parameters to see if we are able to recover the true values. 

We simulate the data from stochastic cusp catastrophe model subject to time-varying volatility as
\begin{eqnarray}
r_t&=& \sigma_ty_t \\
d\sigma_t^2&=&\kappa(\omega-\sigma_t^2)dt+\gamma dW_{t,1},\\
dy_t&=&(\alpha_t+\beta_ty_t-y_t^3)dt+dW_{t,2}
\end{eqnarray}
where $dW_{t,1}$ and $dW_{t,2}$ are standard Brownian motions with zero correlation, $\kappa=5$, $\omega=0.04$ and $\gamma=0.5$. The volatility parameters satisfy Feller's condition $2\kappa\omega\ge\gamma^2$, which keeps the volatility process away from the zero boundary.  We set the parameters to values which are reasonable for a stock price, as in \cite{zhang2005}.
\begin{figure}[t!]
\centering
\includegraphics[scale=0.39]{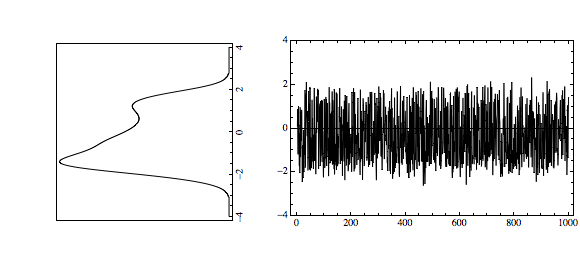}
\includegraphics[scale=0.39]{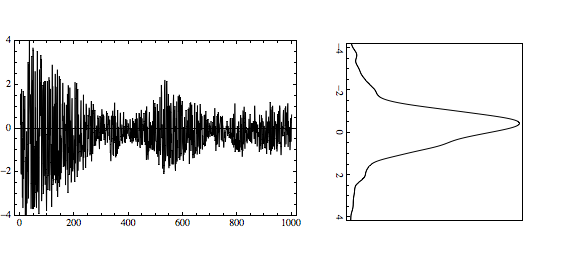}
\caption{An example of simulated time series where cusp surface is subject to noise only, $y_t$, and noise together with volatility, $r_t$. (b) simulated returns $y_t$ (a) kernel density estimate of $y_t$ (c) simulated returns $r_t$ contamined with  volatility (d) kernel density estimate of $r_t$.}
\label{fig:simulatedprice}
\end{figure}

In the cusp equation, we use two independent variables
\begin{eqnarray}
\alpha_t&=&\alpha_0+\alpha_1x_{t,1}+\alpha_2x_{t,2}\\
\beta_t&=&\beta_0+\beta_1x_{t,1}+\beta_2x_{t,2}
\end{eqnarray}
with coefficients $\alpha_2=\beta_1=0$, so $x_1$ drives the assymetry side and $x_2$ drives the bifurcation side of the model solely. Parameters are set as $\alpha_0=-2$, $\alpha_1=3$, $\beta_0=-1$ and $\beta_2=4$.

\begin{figure}[t!]
\centering
\includegraphics[scale=0.4]{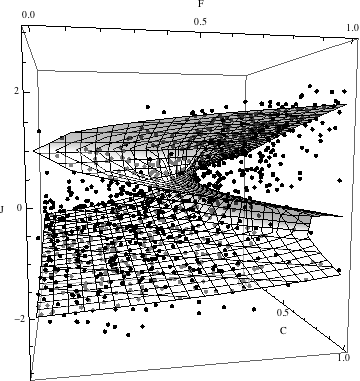}
\includegraphics[scale=0.4]{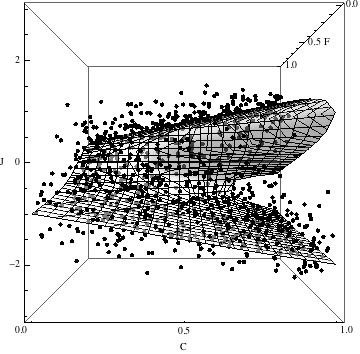}\\
\includegraphics[scale=0.4]{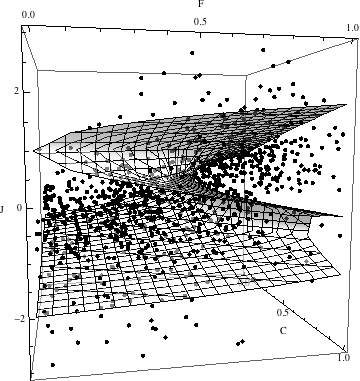}
\includegraphics[scale=0.4]{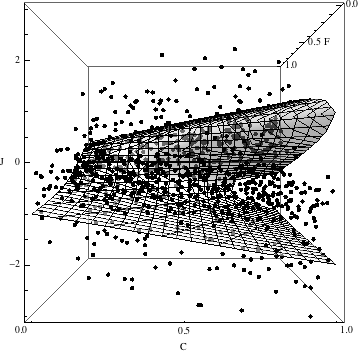}
\caption{An example of simulated data where cusp surface is subject to noise only, $y_t$, and noise with volatility, $r_t$. Parts (a-b) show the cusp deterministic pleat and simulated $\{x_1,x_2,y_t\}$ from 2 different viewpoints, parts (c-d) show the cusp deterministic pleat and simulated $\{x_1,x_2,r_t\}$ from 2 different viewpoints.}
\label{fig:simulatedcusp}
\end{figure}

In the simulations, we are interested to see how the cusp catastrophe model performs under the time-varying volatility. Thus we estimate the coefficients on the processes $y_t=r_t/\sigma_t$ and $r_t$. Figure \ref{fig:simulatedprice} shows one realization of the simulated returns $y_t$ and $r_t$. While  $y_t$ is cusp catastrophe subject to noise, $r_t$ is subject to time-varying volatility as well. It is noticeable how time-varying volatility causes the shift from bimodal density to unimodal. More illustrative is Figure \ref{fig:simulatedcusp}, which shows cusp catastrophe surface of both processes. While the solution from deterministic cusp catastrophe is contamined with noise in the first case, the volatility process in the second case makes it much more difficult to recognize the two states of the system in the bifurcation area. This causes difficulties in recovering true parameters.

Table \ref{tab:simulations} shows the results of simulation. We simulate the processes 100 times and report the mean and standard deviations from the mean\footnote{The distribution of the parameters is Gaussian in both cases which makes it possible to compare the results within mean and standard deviation.}. The true parameters are easily recovered in the simulations from the $y_t$ when cusp catastrophe is subject to noise only as the mean values are statistically indistinguishable from the true simulated values $\alpha_0=-2$, $\alpha_1=3$, $\beta_0=-1$ and $\beta_2=4$. The fits are reasonable as it explains about 60\% of the data variation in the noisy environment. Moreover, in the cusp model, we first estimate the full set of parameters $(\alpha_0,\alpha_1,\alpha_2,\beta_0,\beta_1,\beta_2)$, and second we restrict parameters $\alpha_2=\beta_1=0$. Estimation easily recovers the true parameters in both cases when in unrestricted case estimates $\alpha_2=\beta_1=0$ and fits are statistically the same. In comparison, both cusp models perform much better than logistic and linear models which was expected. It is also interesting to note that $\omega_0=0$ and $\omega_1=1$ which means that the observed data are the true data and no transformation is needed. These results are important as they confirm that estimation of stochastic cusp catastrophe model is valid and it can be used to quantitatively apply the theory on the data. 

\begin{table}[!ht]
\tiny
\centering				
\begin{tabular}{lrrrrrrrrrrr}	
\toprule			
 & \multicolumn{4}{c}{(a) Estimates using $y_t=r_t /\sigma_{t}$ }   &  & \multicolumn{4}{c}{(b) Estimates using $r_t$}  \\  
 \cmidrule{2-5}  \cmidrule{7-10}
 & \multicolumn{2}{c}{Cusp}  &   Linear & Logistic &  & \multicolumn{2}{c}{Cusp}  &  Linear & Logistic\\ 
    \cmidrule{2-5}  \cmidrule{7-10}
 & \multicolumn{1}{c}{Unrestricted}   & \multicolumn{1}{c}{Restricted}   & & &  & \multicolumn{1}{c}{Unrestricted} &  \multicolumn{1}{c}{Restricted} & \\ 
& \\
$\alpha_0$     & -2.059 (0.143)  & -2.066 (0.141)  & & & & -1.614 (3.352)  & -3.111 (0.531)\\ 
$\alpha_1$  & 3.083 (0.195)   & 3.084 (0.203)   & & & & 4.355 (1.382)   & 4.540 (0.690) \\ 
$\alpha_2$  & -0.009 (0.118)  &                 & & & & -1.126 (2.121)  &               \\ 
$\beta_0$     & -1.016 (0.237)  & -1.018 (0.187)  & & & & -7.088 (3.287)  & -5.428 (2.420)\\ 
$\beta_1$  & 0.001 (0.319)   &                 & & & & 2.394 (1.806)   &               \\ 
$\beta_2$  & 4.003 (0.232)   & 4.006 (0.223)   & & & & 5.821 (1.430)   & 5.544 (1.293) \\ 
$\omega_0$    & -0.005 (0.035)  & -0.006 (0.025)  & & & & 0.163 (0.305)   & -0.038 (0.053)\\ 
$\omega_1$   & 1.003 (0.021)   & 1.003 (0.021)   & & & & 0.539 (0.155)   & 0.552 (0.154) \\
 \cmidrule{2-5}  \cmidrule{7-10}
$R^2$ & 0.609 (0.022) & 0.608 (0.023) & 0.397 (0.025) & 0.462 (0.027) &  & 0.378 (0.043) & 0.379 (0.044) & 0.345 (0.034) & 0.401 (0.038)\\ 
LL    & -888.6 (24.3) & -889.4 (24.1) & -1323.3 (21.4) & -1266.4 (23.5) &  & -1109.4 (52.7) & -1117.6 (57.6) & -1471.7 (212.6) & -1426.5 (211.4)\\ 
AIC       & 1793.1 (48.5) & 1790.9 (48.3) & 2654.7 (42.7) & 2546.9 (47.0) &  & 2234.8 (105.5) & 2247.3 (115.3) & 2951.3 (425.1) & 2867.0 (422.9)\\ 
BIC       & 1832.4 (48.5) & 1820.3 (48.3) & 2674.3 (42.7) & 2581.2 (47.0) &  & 2274.0 (105.5) & 2276.7 (115.3) & 2970.9 (425.1) & 2901.4 (422.9) \\
\bottomrule	
\end{tabular}	
\caption{\scriptsize{Simulation results (a) according to stochastic cusp catastrophe model (b) according to stochastic cusp catastrophe model with process in volatility. The total sample based on 100 random simulations is used and sample means as well as standard deviations (in parentheses) for each value are reported. All figures rounded to one or three decimal digits.}}
\label{tab:simulations}
\end{table}

Results of estimation on the $r_t$ process, which is subject to time-varying volatility reveal that addition of the volatility process makes it difficult for maximum likelihood estimation to recover the true parameters. Variance of estimated parameters is very large and the mean values are far away from the true simulated values. Moreover, fits are statistically weaker as it explains no more than 38\% of variance of the data. Also it is interesting to note that logistic fit and linear fit are much closer to the cusp fit.

In conclusion, the simulation results are important as they reveal that time-varying volatility in  cusp catastrophe model destroys the ability of MLE estimator to recover the cusp potential.


\section{Empirical modeling of stock market crashes}

\begin{figure}[t!]
\centering
\includegraphics[scale=0.35]{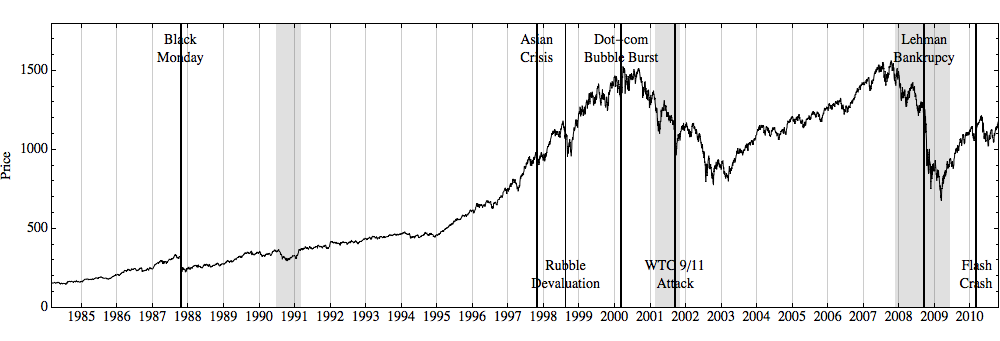}
\caption{S\&P 500 prices data. Figure highlights several important recession periods as grey periods and crashes events as black lines. Periods are closer described in the text.}
\label{fig:prices}
\end{figure}

Armed with the results from simulations, we move to the estimation of cusp catastrophe model on the real-word data from stock markets. For the analysis, we use long time span of S\&P 500, broad U.S. stock market index covering almost 27 years from February 24, 1984 until November 17, 2010. Figure \ref{fig:prices} plots the prices and depicts several recessions and crisis periods. According to the National Bureau of Economic Research (NBER), there were three U.S. recessions in the periods of July 1990 -- March 1991, March 2001 -- Nov 2001 and Dec 2007 -- June 2009. These recessions are depicted as grey periods. Black lines depict one day crashes associated with large price drops. Namely, Black Monday 1987 (October 19, 1987), Assian crisis crash (October 27, 1997), Ruble Devaluation 1998 (August 17, 1998), Dot-com Bubble Burst (March 10, 2000), World Trade Center Attacks (September 11, 2001), Lehman Brothers Holdings Bankruptcy (September 15, 2008), and lastly Flash Crash (March 6, 2010). Technically, largest one day drops in the studied period occurred on October 19, 1987, October 26, 1987, September 29, 2008, October 9, 2008, October 15, 2008, December 1, 2008 recording declines of 20.47\%, 8.28\%, 8.79\%, 7.62\%, 9.03\%, 8.93\%, respectively.

Let us look closer at the crashes depicted by Figure \ref{fig:prices} and discuss their nature. Term Black Monday refers to Monday October 19, 1987 when stock markets around the world from Hong Kong through Europe and U.S. crashed in a very short time and recorded the largest one day drop in history. After this unexpected severe event, many predicted the most troubled years since the 1930s. However, stock markets gained the losses back and closed the year in positive numbers. Until now there has not been consensual opinion on the cause of the crash. Many potential causes include program trading, overvaluation, illiquidity, or market psychology. Thus this crash seems to have an endogenous cause. After several successive years, stock markets did not record large shocks until 1996 when Asian financial crisis driven by investors deserting emerging overheated Asian shares resulted in October 27, 1997 mini crash of U.S. markets. In a successive year, the Russian government devaluated the rubble, defaulted on domestic debt and declared a moratorium on payment to foreign creditors. This caused another international crash on August 17, 1998. These two shocks are believed to be exogenous to the U.S. stock markets. During the period of 1997 -- 2000, the so-called dot-com bubble emerged when a group of internet-based companies entered the markets and attracted many investors confident in their profits overlooking the companies' fundamental values. This resulted in a collapse, or bubble burst during 2000 -- 2001. Another exogenous shock has been brought to stock markets in the year 2001 when World Trade Centre (WTC) was attacked and destroyed. While markets recorded sudden drop, this should not be attributed to internal forces of the markets. The recent financial crisis of 2007 -- 2008 also known as a global financial crisis emerged from bursting of the U.S. housing bubble which peaked in 2006. In a series of days in September and October 2008, stock markets have seen large successive declines. Many believe that this crash has been driven mainly by the housing markets, but there is no consensus about the real causes. Finally, our studied period also covers the May 6, 2010 Flash Crash also known as The Crash of 2:45 in which the Dow Jones Industrial Average index plunged about 1000 points (9\%) only to recover the losses within a few minutes. It was a biggest intraday drop in a history and one of the main possible causes might be impact of high frequency traders or large directional bets. 

Cusp catastrophe theory in terms of \cite{zeeman1974}'s hypotheses proposes to model the crashes as an endogenous events driven by speculative money. Employing our two-step estimation method, we will estimate the cusp model to quantitatively test the theory on period covering all these crashes and see if the theory can explain them using our data. An interesting discussion may stem from studying the causality between volatility and crashes. While \cite{Levy2008} provides a modeling approach for increasing volatility before the crash event, in our approach the crashes are driven endogenously by speculative money, thus the sudden discontinuities are not connected to volatility.

\subsection{Data description}

For our two-step estimation procedure, we need two sets of the data. First set consists of high frequency data and it is used to estimate volatility of returns. Second set consists of a data about sentiment. Let us describe both datasets used. For the realized volatility estimation, we use the S\&P 500 futures traded on the Chicago Mercantile Exchange (CME)\footnote{The data were provided by Tick Data, Inc.}. The sample period extends from February 24, 1984 until November 17, 2010. Although after the introduction of CME Globex(R) electronic trading platform on Monday December 18, 2006, CME started to offer nearly continuous trading, we restrict the analysis on the intraday returns with 5-minute frequency within the business hours of the New York Stock Exchange (NYSE) as the most liquidity of S\&P 500 futures comes from the period when U.S. markets are open. We eliminate transactions executed on Saturdays and Sundays, US federal holidays, December 24 to 26, and December 31 to January 2, because of the low activity on these days, which could lead to estimation bias. 
\begin{figure}[t!]
\centering
\includegraphics[scale=0.35]{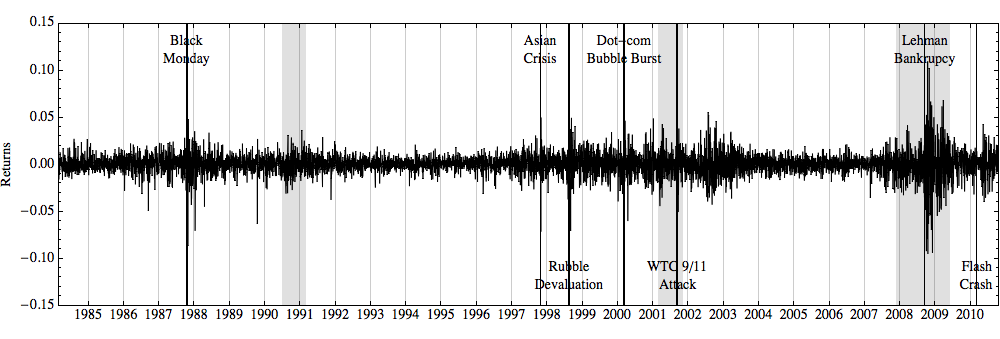}
\includegraphics[scale=0.35]{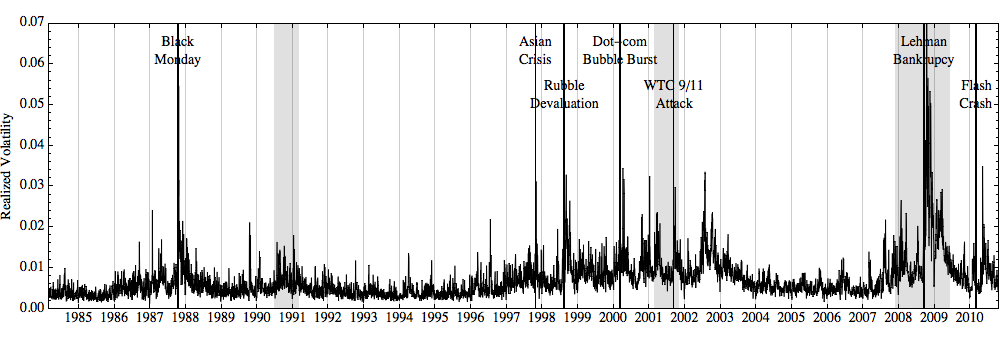}
\includegraphics[scale=0.35]{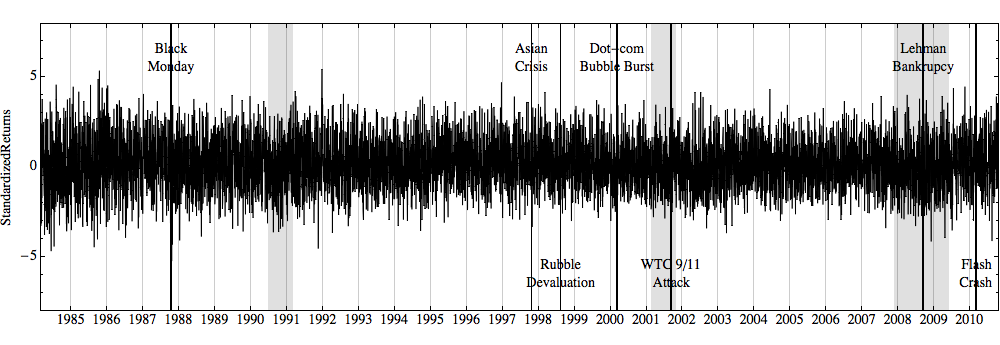}
\caption{S\&P 500 (a) returns $r_t$, (b) realized volatility $RV_t$ and (c) standardized returns $r_t RV_t^{-1/2}$ data. Figure highlights several important recession periods as grey periods and crashes events as black lines. Periods are closer described in the text.}
\label{fig:returns}
\end{figure}

Using the realized volatility estimator, we then measure the volatility of stock market returns as a sum of squared 5-minute intraday returns. In this way, we obtain 6739 daily volatility estimates. Figure \ref{fig:returns} shows the estimated volatility \ref{fig:returns}(b) together with daily returns \ref{fig:returns}(a). It can be immediately noticed that the volatility of the S\&P 500 is strongly time-varying over the very long period. 

For the state (behavioral) variable of the cusp model, we use S\&P 500 daily returns standardized by the estimated daily realized volatility according to Eq. \ref{eq:rv}. By standardization, we obtain stationary data depicted by the Figure \ref{fig:returns}(c).

In choosing control variables, we follow the success of our previous application in \cite{barunik2009} where we have compared several measures of control variables and we have shown that fundamentalists, or the asymmetry side of the market, are best described by the ratio of advancing stocks volume and declining stock volume and chartists, or the bifurcation side of the model, are best described by the OEX put/call ratio\footnote{The data were provided by Pinnacle Data Corp.}. Variables related to the trading volumes are generally considered as a good measure of trading activity of large funds and institutional investors and thus representing the fundamental side of the market. Therefore, the ratio of advancing stock and declining stock volume should contribute mainly to the asymmetry side of the model. On the other hand, the activity of market speculators and technical traders should be well captured by the OEX put/call ratio as it is the ratio of daily put volume and call volume of the options with the underlying Standard and Poors' index. Financial options are widely used and the most popular instruments for speculative purposes and therefore serve as a good measure of speculative money. Thus they should represent internal forces leading the market to the bifurcation within cusp catastrophe model. All in all, we suppose the OEX put/call options ratio to contribute mainly on the bifurcation side of the model. Moreover, we use the third control variable, which is the daily change of total trading volume, as a driver for both fundamental and speculative money in the market.  It is generally related to the continuos fundamental trading activity but it may also reflect the elevated speculative activity on the market well. Therefore, we expect this variable not only to help the regression on the asymmetry side, but also on the bifurcation side. The time span of all these data matches the time span of S\&P 500 returns, i. e. February 24, 1984 until November 17, 2010. The descriptive statistics of all the data used are saved in the Table \ref{tab:stats} in Appendix \ref{app:descr}.

\subsection{Full sample static estimates}

In the estimation, we primarily aim at testing whether cusp catastrophe model is able to describe the stock market data well in the time-varying volatility environment and thus that stock markets show signs of bifurcations. In doing so, we follow the statistical testing described earlier in the text. 

\begin{table}[tb!]
\footnotesize
\centering				
\begin{tabular}{lrrrrrrrrrrrrrrr}	
\toprule			
 & \multicolumn{6}{c}{(a) Estimates using $r_t \widehat{RV}_{t}^{-1/2}$}   &  & \multicolumn{6}{c}{(b) Estimates using $r_t$}  \\  
 \cmidrule{2-7}  \cmidrule{9-14}
 & \multicolumn{4}{c}{Cusp}  &   Linear & Logistic &  & \multicolumn{4}{c}{Cusp}  &  Linear & Logistic\\ 
 \cmidrule{2-7}  \cmidrule{9-14}
 & \multicolumn{2}{c}{Unrestricted}   & \multicolumn{2}{c}{Restricted}   & & &  & \multicolumn{2}{c}{Unrestricted} &  \multicolumn{2}{c}{Restricted} & \\ 
& \\
$\alpha_0$ & -2.378 &*** & 1.333 &*** &  &  &  & 5.008 &*** & 5.000 & *** &  & \\ 
$\alpha_1$ & 4.974 &*** & 3.369  &*** &  &  &  & 4.426  &*** & 2.730 & *** &  & \\ 
$\alpha_2$ & 0.062 &*** &  &  & &   &  & 0.771  &*** && &   & \\ 
$\alpha_3$ & 0.300 &*** & -0.064  &*** &  &  &  & -0.463  &*** & -0.534 & *** &  & \\ 
$\beta_0$ & -4.654 &*** & -1.334  &*** &  &  &  & -5.011  &*** & -5.000 & *** &  & \\ 
$\beta_1$ & -5.011 &*** &  &  &  &  &  & -1.201  &*** &  &&  & \\ 
$\beta_2$ & 0.139 &*** & -0.054  &*** &  &  &  & -1.138  &*** & -0.208 & *** &  & \\ 
$\beta_3$ & 0.422 &*** & 0.299  &*** &  &  &  & 0.644  &*** & 0.683 & *** &  & \\ 
$\omega_0$ & -0.700 &*** & 0.329  &*** &  &  &  & 0.786 & *** & 0.826 & *** &  & \\ 
$\omega_1$ & 0.492 &*** & 0.905  &*** &  &  &  & 0.407 & *** & 0.402 &*** &  & \\ 
 \cmidrule{2-7}  \cmidrule{9-14}
$R^2$ & \multicolumn{2}{r}{0.800 }& \multicolumn{2}{r}{0.769}   & 0.385 & 0.823 &  &  \multicolumn{2}{r}{0.637}   &  \multicolumn{2}{r}{0.530}   & 0.405 & 0.687\\ 
LL &   \multicolumn{2}{r}{ -4786.161}  & \multicolumn{2}{r}{-4733.071}  & -7634.756 & -5885.983 &  & \multicolumn{2}{r}{ -7728.981}   &  \multicolumn{2}{r}{-8174.635}   & -7811.134 & -5648.799\\ 
AIC & \multicolumn{2}{r}{9592.321}  & \multicolumn{2}{r}{9482.142} &   15279.512 & 11789.970 &  &  \multicolumn{2}{r}{15477.960}   &  \multicolumn{2}{r}{16365.270}  & 15632.270 & 11311.600\\ 
BIC & \multicolumn{2}{r}{9660.209}  & \multicolumn{2}{r}{9533.682} &   15313.456 & 11851.070 &  &  \multicolumn{2}{r}{15545.820}   &  \multicolumn{2}{r}{16419.800}   & 15666.350 & 11359.310 \\
\bottomrule	
\end{tabular}		
\caption{Estimation results on the S\&P 500 stock market data. Full sample extends from February 24, 1984 until November 17, 2010. Left part of the Table presents the estimation results on the normalized returns $r_t \widehat{RV}_{t}^{-1/2}$, right part of the Table presents the results for the original S\&P 500 stock market returns $r_t$.}
\label{tab:estimates1}
\end{table}
Table \ref{tab:estimates1} shows the estimates of the cusp fits. Let us concentrate of the left part of the Table \ref{tab:estimates1}(a), where we fit cusp catastrophe model to the standardized returns $r_t \widehat{RV}_{t}^{-1/2}$. First, we do not make any restrictions and use all the three control variables, thus $\alpha_x=\alpha_0+\alpha_1x_{1,t}+\alpha_2x_{2,t}+\alpha_3x_{3,t}$ and $\beta_x=\beta_0+\beta_1x_{1,t}+\beta_2x_{2,t}+\beta_3x_{3,t}$, where $x_1$ is the ratio of advancing stocks volume and declining stocks volume, $x_2$ OEX put/call options ratio and $x_3$ rate of change of the total volume. State variable is $\tilde{r}_t$, returns normalized with estimated realized volatility. In terms of log likelihood, cusp model describes the data much better than linear regression model. $\omega_1$ coefficient is far away from zero although some degree of transformation of the data is needed. All the other coefficients are strongly significant at 99\% levels. Most importantly, when cusp fit is compared to the logistic fit in terms of AIC and BIC, we can see that the cusp model strongly outperforms the logistic model. 

Our hypothesis is, that the ratio of advancing and declining stock volume will contribute only to the asymmetry side and OEX put/call ratio representing the measure of speculative money in the market will contribute to the bifurcation side of the model. For testing this hypothesis, we set the parameters $\alpha_2=\beta_1=0$ and refer to this as a restricted model. From the Table \ref{tab:estimates1}(a), we can see that log likelihood of the restricted model improves in comparison to the unrestricted one. Also in terms of information criteria, fit further improves. All the parameters are again strongly significant and we can see that they change considerably. This can be attributed to the fact that $x_{1,t}$ seems to contribute strongly to both sides of the market in the unrestricted model.  Although $\beta_2$ coefficient representing the coefficient of speculative money is quite small in comparison to other coefficients, still it is strongly significant. As this is the key coefficient of the model driving the stock market to the bifurcation, we will further investigate its impact in following sections. It is interesting to note that the $\omega_1$ parameter increases very near to one in the restricted model. This means that the observed data are close to the state variable. 

When moving to the right (b) part of the Table \ref{tab:estimates1}, we repeat the same analysis, but this time, we use original $r_t$ returns as the state variable. This means that we would like to compare cusp catastrophe fit on the data with strongly varying volatility. While using the very long time span of the data where the volatility varies considerably, we expect the model to deteriorate. Although the application of cusp catastrophe model to the non-stationary data can be questioned, we bring these estimates for comparison to our modeling approach. We can see an important result here. While linear and logistic models provide the very similar fits in terms of log likelihoods, information criteria and $R^2$, both unrestricted as well as restricted cusp models deteriorate. $\omega_1$ coefficient together with all the other coefficients are still strongly different from zero, but important result is that not only the logistic model describes the data better, but also the presence of bifurcations in the raw return data can not be claimed.

To conclude this section, the results suggest strong evidence that over the long period of almost 27 years, stock markets are better described by cusp catastrophe model. Using our two-step modeling approach we have shown that cusp model fits the data far best, fundamental and bifurcation sides are controlled by the indicators of fundament and speculative money, respectively. In a contrast, when cusp is fit to the original data under strong variation of volatility, model deteriorates. We should note that these results resemble the results from simulation, thus simulation strongly supports our modeling approach as well.

\subsection{Examples of 1987 and 2008 crashes}

\begin{table}[tb!]
\footnotesize
\centering				
\begin{tabular}{lrlrlrrrrlrlrrrr}	
\toprule			
 & \multicolumn{6}{c}{1987}   &  & \multicolumn{6}{c}{2008}  \\  
 \cmidrule{2-7}  \cmidrule{9-14}
 & \multicolumn{4}{c}{Cusp}  &   Linear & Logistic &  & \multicolumn{4}{c}{Cusp}  &  Linear & Logistic\\ 
 \cmidrule{2-7}  \cmidrule{9-14}
 & \multicolumn{2}{c}{Unrestricted}   & \multicolumn{2}{c}{Restricted}   & & &  & \multicolumn{2}{c}{Unrestricted} &  \multicolumn{2}{c}{Restricted} & \\ 
& \\
$\alpha_0$ & -0.970  &* & -0.535 & *** &  &  &  & -3.938&  *** & -0.598 & ***\\ 
$\alpha_1$ & 1.792  &*** & 1.794& *** &  &  &  & 1.793 & *** & 1.798 & ***\\ 
$\alpha_2$ & -0.073  & &    &  & & &  & -1.322  && &   \\ 
$\alpha_3$ & 0.191 &  & 0.029  & &  &  &  & -0.814 &  & 0.005 & ***\\ 
$\beta_0$ & 0.562  &* & 0.322 & * &  &  &  & -0.542 & *** & -1.142 & *\\ 
$\beta_1$ & -0.395  &* &  &  &  &  &  & -1.803 & *** & &&  \\ 
$\beta_2$ & -1.255  &** & -0.731 & * &  &  &  & -0.990 & ** & -0.031 & ***\\ 
$\beta_3$ & 1.648  &*** & 1.547  &*** &  &  &  & -0.803  & & 0.282 & \\ 
$\omega_0$  & 0.453 & *** & 0.771 &*** &  &  &  & -0.637 & *** & 0.680 & ***\\ 
$\omega_1$  & 0.561  &*** & 0.602 & *** &  &  &  & 0.476 & *** & 0.641 & *** \\
 \cmidrule{2-7}  \cmidrule{9-14}
$R^2$ &  \multicolumn{2}{r}{0.855}   &  \multicolumn{2}{r}{0.827} &   0.454 & 0.895 & &    \multicolumn{2}{r}{0.816}  &  \multicolumn{2}{r}{0.858}   & 0.483 & 0.884\\ 
LL &  \multicolumn{2}{r}{-85.971}   &  \multicolumn{2}{r}{-90.659} &   -208.557 & -104.045  & &  \multicolumn{2}{r}{-88.054}  &  \multicolumn{2}{r}{-103.437} & -185.150 & -90.148\\ 
AIC &  \multicolumn{2}{r}{191.943} &    \multicolumn{2}{r}{197.317} &   427.114 & 226.089  & &  \multicolumn{2}{r}{196.108}  &  \multicolumn{2}{r}{222.875}  & 380.300 & 198.296\\ 
BIC &  \multicolumn{2}{r}{220.385} &    \multicolumn{2}{r}{220.071} &   441.335 & 251.687  & &  \multicolumn{2}{r}{224.550}  &  \multicolumn{2}{r}{245.628}  & 394.521 & 223.894 \\
\bottomrule	
\end{tabular}		
\caption{Estimation results of the two distinct periods of the S\&P 500 normalized stock market returns, $r_t \widehat{RV}_{t}^{-1/2}$.}
\label{tab:estimates2}
\end{table}

While the results in the previous section are supportive for cusp catastrophe model, the sample period of almost 27 years might contain many structural changes. Thus we would like to further investigate how the model performs in time. For this, we borrow the two very distinct crashes of 1987 and 2008 and compare the localized cusp fits. The reasons for studying these particular periods are several. These two crashes are distinct in time as there is 21 years difference between them, so we would like to see how the data describe the periods. The stock market crash of 1987 has not yet been explained and many believe that it was an endogenous crash constituting perfect candidate for the cusp model. On the other hand, the 2008 period covers much deeper recession, thus is very different from the 1987. Last important reason is that both periods contain all the largest one day drops, which occurred on October 19, 1987, October 26, 1987, September 29, 2008, October 9, 2008, October 15, 2008, December 1, 2008 recording declines of 20.47\%, 8.28\%, 8.79\%, 7.62\%, 9.03\%, 8.93\%, respectively. In the following estimations, we will restrict ourselves to our newly proposed two-step approach of cusp catastrophe fitting procedure and we utilize the sample covering half year. 

When focusing on the estimation results for the 1987 crash, we can see that both restricted and unrestricted models fit the data much better than linear regression. $\omega_1$ coefficients are significantly different from zero and when the cusp models are compared to the logistic model, they also seem to provide much better fits. Thus cusp catastrophe model explains the data very well and we can conclude that the stock market crash of 1987 has been led by internal forces. This confirms our previous findings in \cite{barunik2009} even though comparison can not be made directly as we have used different sample length in our previous study. When comparing unrestricted and restricted model fits, we can see that they do not differ significantly within log likelihoods, AIC and BIC. Also coefficient estimates are close. The reason is that unrestricted model estimates the $\alpha_2$ coefficient which can not be distinguished from zero and $\beta_1$ coefficient is significant only on 90\% level of significance. Thus $x_1$ proves to drive the fundamentals and $x_2$ speculators. Note that $\beta_2$ coefficient is much larger in its magnitude than on the fit of full sample in the previous section. Interestingly, $x_3$ seems to drive the speculative money in the 1987, but does not help to explain the 2008 behavior.


Data from the 2008 period show different results. While cusp fits are much better in comparison to the linear regression, they can not be statistically distinguished from logistic model. This means that there is very weak evidence of discontinuities in this period. It is an interesting result, as it may suggest that the 2008 large drops were not driven endogenously by stock market participants but exogeneously by the burst of the housing market bubble.




\subsection{Rolling regression estimates}

While 1987 data are explained by cusp catastrophe model very well and 2008 are not, we would like to further investigate how cusp catastrophe fit changes over time. With almost 27 years of all data needed for our two-step method of estimation, we estimate cusp catastrophe model on half year samples rolling with the step of one month. The half year period is reasonable as it is enough data for the sound statistical fit and on the other hand it is not too long period so it can uncover a structural breaks in the data. In the estimation, we again restrict ourselves to our two-step estimation procedure. To keep the results under control, we use final restricted model, where we suppose that $x_1$ controls asymmetry side of the model and $x_2$ controls bifurcation side solely, while $x_3$ contributes to both sides. Thus $\alpha_2=\beta_1=0$. 

\begin{figure}[tb!]
\centering
\includegraphics[scale=0.4]{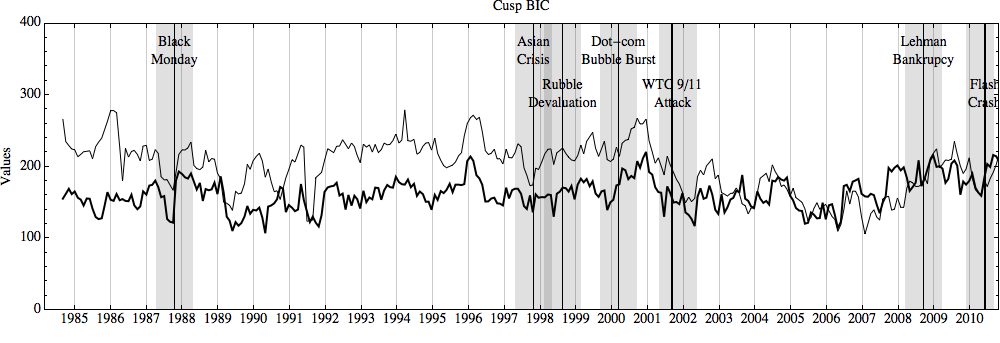}
\caption{Rolling values of BIC information criteria for cusp catastrophe (in bold black) and logistic (in black) models.}
\label{fig:rollingbic}
\end{figure}

\begin{figure}[htb!]
\centering
\includegraphics[scale=0.4]{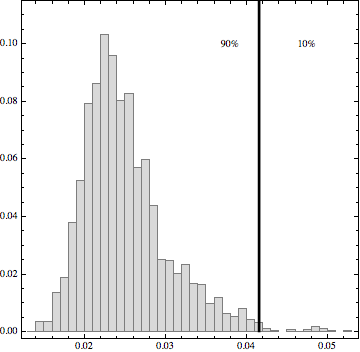}
\caption{Histogram of the dip statistics for bimodality computed for all rolling window periods together with bootstrapped critical value 0.0406 for the 90\% significance level plotted in bold black.}
\label{fig:bimodalitytest}
\end{figure}

Before we move to interpreting the rolling regression results, let us discuss bimodality of the rolling samples. Stochastic catastrophe corresponds to a transition from a unimodal to a bimodal distributions. Thus we need to first test for the bimodality in order to be able to draw any conclusions from our analysis. For this, we borrow the dip test of unimodality developed in \cite{hartigan1985algorithm,hartigan1985dip}. The dip statistic is the maximum difference between the empirical distribution function and the unimodal distribution function and it measures departure of the sample from unimodality. Asymptotically the dip statistics for samples from a unimodal distribution approaches zero and for samples from any multimodal distribution approaches a positive constant. We use bootstrapped critical values for the small rolling sample sizes in order to asses the unimodality. Figure \ref{fig:bimodalitytest} shows the histogram of all dip statistics together with its bootstrapped critical value 0.0406 for the 90\% significance level. The results suggest that unimodality is rejected at 90\% significance level at several periods, while for most of the periods, unimodality can not be rejected. Thus we observe transition from unimodal to bimodal (or possibly multimodal) distributions during the studied period several times. 

\begin{figure}[htb!]
\centering
\includegraphics[scale=0.34]{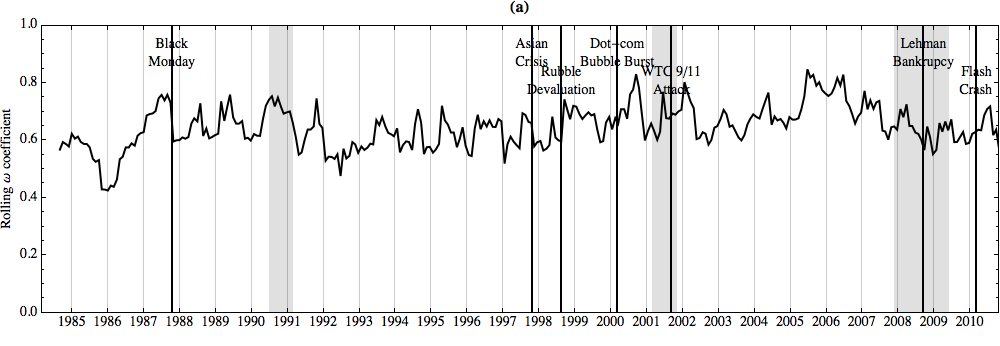}
\includegraphics[scale=0.34]{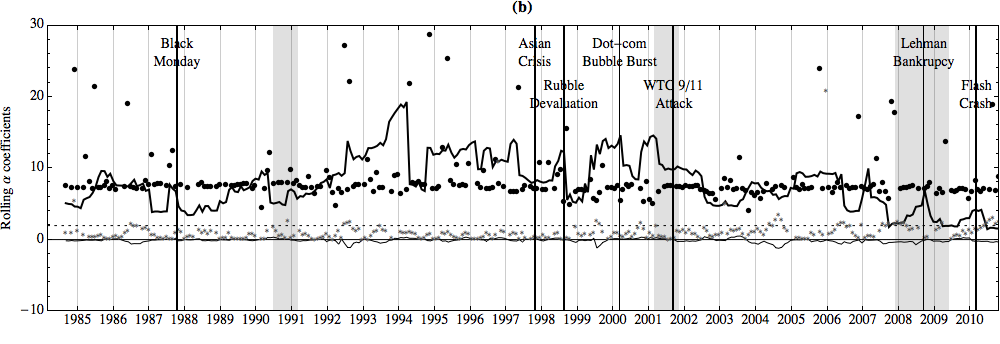}
\includegraphics[scale=0.34]{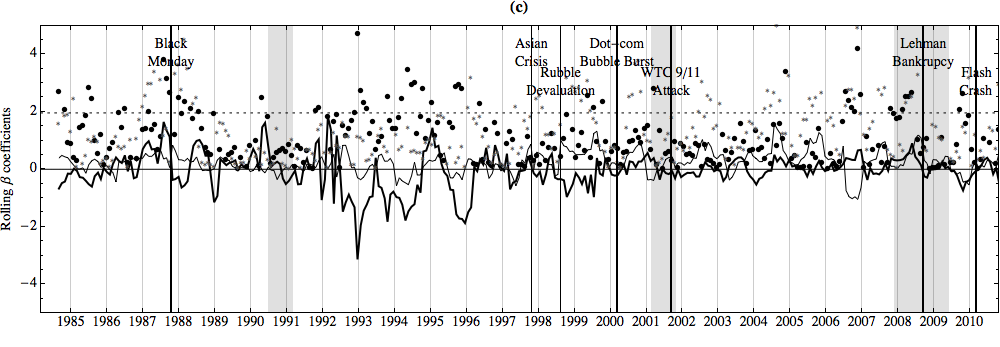}
\caption{Rolling coefficients with their $|$z-values$|$. (a) estimated $\omega_1$ coefficient values, (b) estimated values of asymmetry coefficients, $\alpha_1$ in bold black, $\alpha_3$ in black. $|$z-values$|$ related to both coefficient estimates are as $\bullet$ and $*$, respectively. (c) estimated values of bifurcation coefficients, $\beta_2$ in bold black, $\beta_3$ in black. $|$z-values$|$ related to both coefficient estimates are as $\bullet$ and $*$, respectively. Plots (b) and (c) also contain 95\% reference z-value as a dashed black line. }
\label{fig:rollingomega}
\end{figure}

Encouraged by the knowledge that the bifurcations might be present in our dataset, we move to rolling cusp results. Figure \ref{fig:rollingomega} shows the rolling coefficient estimates together with its significance. $\omega_1$ is significantly different from zero in all periods. $\alpha_1$ coefficient is strongly significant over the whole period, while it becomes lower in its magnitude during the last years. Thus ratio of advancing and declining volumes is a good measure for fundamentalists driving the asymmetry part of the model. Much more important is the $\beta_2$ parameter, which drives the bifurcations in the model. We can notice that while in the period until the year 1996 it has been significant and its value has changed over time considerably, in the period after the year 1996 it can not be distinguished from zero statistically (except for some periods). This is very interesting result as it shows how OEX put/call ratio was a good measure of speculative money in the market and controlled the bifurcation side of the model. During the first period, it was driving the stock market into the bifurcations, while in the second period, the market was rather stable under the model. Only in the very last few years in the recent 2008 recessions parameter started to play a role in the model again. Still, its contribution is relatively small. 

This is confirmed also when we compare the cusp model to the logistic one. Figure \ref{fig:rollingbic} compares the Bayesian Information Criteria of the two models and we can see that cusp catastrophe model was much better fit of the data up to year 2003, while for the upcoming years of 2003-2009 (roughly) it can not be distinguished from the logistic model, or logistic model outperforms the cusp strongly. In the last period after 2009 and before the Flash crash, cusp again explains the data better, but the difference is not so strong as in the before-2003 period. This result in fact shows that while before 2003 stock markets were showing marks of bifurcation behavior according to the cusp model, later on after the year 2003 in the period of stable growth when stock market participants believed that stock markets are stable, they do not show marks of the bifurcation behavior any more.

To conclude this section, we have found that in spite of the fact that we model volatility in the first step, stock markets show marks of bistability over several crisis periods.

\section{Conclusion}

In this paper, we contribute to the stock market crashes modeling literature and quantitative application of stochastic cusp catastrophe theory. We develop a two-step estimation procedure and estimate cusp catastrophe model under the time-varying stock market volatility. This allows us to test the \cite{zeeman1974}'s qualitative hypotheses about cusp catastrophe and bring new empirical results to our previous work in this area \citep{barunik2009}.

In the empirical testing, we use a unique high-frequency and sentiment data about U.S. stock market evolution covering almost 27 years. The results suggest evidence that over the long period, stock markets were better described by stochastic cusp catastrophe model. Using our two-step modeling approach we have shown that cusp model fits the data far best, fundamental and bifurcation sides are controlled by the indicators of fundament and speculative money, respectively. In a contrast, when cusp is fit to the original data under strong variation of volatility, model deteriorates. We should note that these results resemble the results from a Monte Carlo study we run, thus simulation strongly supports our analysis.  Further, we develop a rolling estimation and find that while until the year 2003 cusp catastrophe model explains the data well, this changes in the period of stable growth 2003--2008. 

In conclusion, we have found that in spite of the fact that we model volatility in the first step, stock markets show marks of bistability during several crisis periods.


%
%

\bibliography{RCuspreferences}
\bibliographystyle{chicago}

\newpage
\section*{Appendix: Descriptive statistics}
\label{app:descr}

\begin{table}[!h]
\footnotesize 
\begin{center}
\begin{tabular}{lrrrrrr}

\toprule
    & $r_t$ & $RV_t$ & $r_t RV_t^{-1/2}$ & $x_1$ & $x_2$ & $x_3$ \\
\hline
Mean            & 0.00030 & 0.00697 & 0.15535 & 1.67051 & 1.16636 & 0.02029\\
Median         & 0.00059 & 0.00571 & 0.11005 & 1.12730 & 1.11000 & 0.00301\\
Std. Dev.      & 0.01175 & 0.00478 & 1.42534 & 2.21562 & 0.36245 & 0.21405\\
Skewness    & -1.32728 & 3.55338 & 0.13660 & 6.71651 & 1.37368 & 2.38477\\
Ex. Kurtosis & 29.44420 & 23.96040 & -0.02713 & 76.35370 & 4.87145 & 17.36620\\
Minimum      & -0.22887 & 0.00122 & -5.21961 & 0.00187 & 0.30000 & -0.76040\\
Maximum     & 0.10957 & 0.07605 & 5.45738 & 44.06780 & 4.56000 & 2.19614\\
\bottomrule

\end{tabular}
\caption{\scriptsize{Descriptive Statistics of the data. Sample period extends from February 24, 1984 until November 17, 2010. The S\&P 500 stock market returns $r_t$, realized volatility $RV_t$, daily returns normalized by the realized volatility $r_t RV_t^{-1/2}$ and data for independent variables $\{x_1,x_2,x_3\}$ are ratio of advancing stocks volume and declining stock volume, OEX put/call options and change of total volume, respectively.}}
\label{tab:stats}
\end{center}
\end{table}

\newpage
\section*{Appendix: Rolling regression estimates}



\end{document}